\documentclass[aps,twocolumn,showpacs,preprintnumbers,nofootinbib]{revtex4-1}

\usepackage{amsmath,amssymb,bm}
\usepackage[dvips]{graphicx}
\usepackage{hyperref}
\usepackage{yfonts}
\usepackage{color}
\newcommand{\beq}{\begin{eqnarray}}
\newcommand{\eeq}{\end{eqnarray}}
\newcommand{\U}{\text{U}}
\renewcommand\d{\partial}
\begin{document}

\title{Scaling laws in chiral hydrodynamic turbulence}

\author{Naoki Yamamoto}
\affiliation{Department of Physics, Keio University,
Yokohama 223-8522, Japan}

\begin{abstract}
We study the turbulent regime of chiral (magneto)hydrodynamics for charged and neutral 
matter with chirality imbalance. We find that the chiral magnetohydrodynamics for charged 
plasmas possesses a unique scaling symmetry, only without fluid helicity under the local 
charge neutrality. We also find a different type of unique scaling symmetry in the chiral 
hydrodynamics for neutral matter with fluid helicity in the inertial range. We show that these 
symmetries dictate the self-similar inverse cascade of the magnetic and kinetic energies. 
Our results imply the possible inverse energy cascade in core-collapse supernovae due to the 
chiral transport of neutrinos.
\end{abstract}
\maketitle

\section{Introduction}

Recently, chiral transport phenomena related to quantum anomalies \cite{Adler,BellJackiw} 
have attracted much attention  both theoretically and experimentally, especially in heavy ion 
physics \cite{Kharzeev:2015znc} and a new type of materials named the Weyl (semi)metals 
\cite{Hosur:2013kxa}. Of particular interest is the possible observation of the current along 
a magnetic field in the presence of the chirality imbalance, called the chiral magnetic effect 
(CME) \cite{Vilenkin:1980fu,Nielsen:1983rb,Alekseev:1998ds,Fukushima:2008xe}. 
Such unusual transport phenomena could potentially lead to some physical consequences 
in other relativistic systems, such as the electroweak plasmas in the early Universe 
\cite{Joyce:1997uy,Boyarsky:2011uy}, electromagnetic plasmas in neutron stars 
\cite{Charbonneau:2009ax} and core-collapse supernovae \cite{Ohnishi:2014uea,Yamamoto:2015gzz}, 
and so on. 

To describe these chiral transport phenomena in nonequilibrium situations, hydrodynamics 
and kinetic theory have been reformulated, which are now referred to as the chiral 
(or anomalous) hydrodynamics \cite{Son:2009tf} and chiral kinetic theory 
\cite{Son:2012wh,Stephanov:2012ki,Chen:2012ca}, respectively. However, the evolutions 
of chiral matter when \emph{nonlinear} effects of the fluid velocity and/or \emph{dynamical} 
electromagnetic fields become important have not been fully understood so far; see 
Refs.~\cite{Giovannini:2013oga,Boyarsky:2015faa} for recent related works. For analytical 
and numerical analyses of chiral hydrodynamics in {\it external} electromagnetic fields, 
see, e.g., Refs.~\cite{Yamamoto:2015ria, Abbasi:2015saa} 
and Ref.~\cite{Hongo:2013cqa}, respectively.

In this paper, we study the generic properties of the chiral (magneto)hydrodynamics 
describing the evolutions of charged and neutral matter at finite chiral chemical potential 
$\mu_5$ and finite temperature $T$. We find that the chiral magnetohydrodynamics 
(ChMHD), together with the chiral anomaly relation, possesses a \emph{unique} scaling 
symmetry for $\mu_5 \ll T$ under the local charge neutrality without fluid helicity. We also 
find a different type of unique scaling symmetry in the chiral hydrodynamics for neutral 
matter at finite chemical potential $\mu$ in the presence of fluid helicity in the so-called 
``inertial range" where dissipation is negligible. We stress that the presence of quantum 
anomalies and chiral transport phenomena is important for these scaling symmetries.

From these scaling symmetries, we derive the self-similar scaling laws of the magnetic and 
kinetic energies [see Eqs.~(\ref{E_B-psi}) and (\ref{E_v-psi})] and the scaling laws of the 
magnetic and kinetic correlation lengths in chiral matter 
[see Eqs.~(\ref{xi_B_t}), (\ref{xi_v_t}), and (\ref{xi_v_t_n})]. 
These results show that the inverse energy cascade---the process that transfers the 
energy from small to larger scales---occurs in the turbulent regime of both ChMHD and 
neutral chiral hydrodynamics under the conditions above. In particular, it implies that the 
chiral transport of neutrinos \cite{Yamamoto:2015gzz} neglected so far may lead to the 
inverse energy cascade in core-collapse supernovae, instead of the direct energy cascade 
observed in the conventional neutrino transport theory \cite{Hanke:2011jf}. This qualitative 
modification may be potentially important to understanding the origin of supernova explosions.

The paper is organized as follows: In Sec.~\ref{sec:ChMHD}, after reviewing the ChMHD 
equations, we discuss its applicability and the conservation laws.%
\footnote{To our knowledge, the applicability of the ChMHD has not been appreciated 
earlier, except for Refs.~\cite{Akamatsu:2013pjd, Akamatsu:2014yza}. The regime of 
applicability will be essential for determining the scaling symmetry of the ChMHD below.}
We then study the scaling symmetry of the ChMHD and its physical consequences. In 
Sec.~\ref{sec:ChHD} we discuss the scaling symmetry of the neutral chiral hydrodynamics
and its physical applications. Section \ref{sec:discussion} is devoted to summary and 
discussion.

In the following, we use the natural units $\hbar = c = 1$.

\section{Chiral magnetohydrodynamics for charged plasmas}
\label{sec:ChMHD}
Let us first consider the ChMHD for plasmas of a Dirac fermion at finite chiral chemical 
potential $\mu_5 \equiv (\mu_{\rm R}-\mu_{\rm L})/2$. We will be interested in the time 
scale larger than $1/\sigma$ (with $\sigma$ being the electrical conductivity), during which 
the electric charge diffuses immediately. Then, we can assume the local charge neutrality, 
$n=0$ or $\mu \equiv (\mu_{\rm R}+\mu_{\rm L})/2=0$. 
On the other hand, the chiral charge $n_5$ can be generally finite in this regime.
We will see in Sec.~\ref{sec:applicability} that there is actually some constraint for $\mu_5$
to treat it as a hydrodynamic (slow) variable.

\subsection{Hydrodynamic equations}
\label{sec:equation}
The ChMHD equations are obtained by promoting external electromagnetic fields 
in the chiral hydrodynamic equations of Ref.~\cite{Son:2009tf} to dynamical ones.
Note here that the chiral charge $n_5$ also evolves in time and space, according to 
the chiral anomaly relation in the presence of electromagnetic fields [see Eq.~(\ref{chiral})], 
and it should be regarded as a {\it dynamical} variable $n_5(t,{\bm x})$ as well.
As the electromagnetic fields and the chiral charge vary much faster (and at a shorter 
length scale) than $T$, we assume that $T$ is static and homogeneous in the regime 
of our interest.
Keeping the main applications of the ChMHD to chiral plasmas in the early Universe and 
astrophysical systems in mind, we also assume the bulk fluid velocity to be nonrelativistic, 
$v \equiv |{\bm v}| \ll 1$, and we only retain the terms to the leading order in $v$ below. 
We will discuss the case with (ultra)relativistic bulk fluid velocity $v \sim 1$ in Sec.~\ref{sec:discussion}, 
which may be relevant to quark-gluon plasmas in heavy ion collisions.

The hydrodynamic equations in the Landau-Lifshitz frame are given by 
\cite{Son:2009tf,Yamamoto:2015gzz}%
\footnote{Precisely speaking, we have an additional charge density due to the CME 
as $\Delta n = \kappa_B \mu_5 {\bm v} \cdot {\bm B}$ \cite{Yamamoto:2015gzz}.
However, it can be shown that its contribution to the right-hand side of Eq.~(\ref{EM}), 
expressed by $\Delta n {\bm E}$, is negligibly small compared with the term 
${\bm j} \times {\bm B}$ under the condition (\ref{condition}) derived below.}
\begin{gather}
\label{EM}
\d_{\mu}T^{\mu\nu} = F^{\nu\lambda}j_{\lambda}, \\
\label{vector}
{\bm \nabla}\cdot {\bm j} = 0, \\
\label{chiral}
\d_t (n_5 + \kappa {\bm v} \cdot {\bm \omega}) + {\bm \nabla}\cdot {\bm j}_5 = C{\bm E}\cdot{\bm B},
\end{gather}
together with Maxwell's equations,
\beq
\label{Maxwell}
\d_{\nu} F^{\nu \mu} = j^{\mu}.
\eeq
Here the energy-momentum tensor $T^{\mu \nu}$, and the vector and axial currents, 
${\bm j}$ and ${\bm j}_5$, are given by \cite{Son:2009tf}%
\footnote{We denote the anomalous transport coefficients by $\kappa$ and $\kappa_B$ 
instead of $\xi$ and $\xi_B$ in Refs.~\cite{Son:2009tf,Yamamoto:2015gzz}, as $\xi$ will
be used for the correlation length later.}
\begin{gather}
T^{\mu \nu}=(\epsilon+P)u^{\mu}u^{\nu}-Pg^{\mu \nu}+\tau^{\mu \nu}, \\
\label{j}
{\bm j}=\sigma({\bm E}+{\bm v}\times {\bm B}) + \kappa_B {\bm B}, \\
\label{j_5}
{\bm j}_5=n_5{\bm v} + \kappa {\bm \omega},
\end{gather} 
where $\epsilon$ is the energy density, $P$ is the pressure, $\sigma$ is the electrical 
conductivity, ${\bm \omega} = {\bm \nabla} \times {\bm v}$ is the vorticity, and $\tau^{\mu \nu}$ 
expresses the dissipative effects like viscosity. Equation (\ref{chiral}) expresses the violation 
of the axial current conservation by the chiral anomaly \cite{Adler,BellJackiw} and the mixed 
gauge-gravitational anomaly \cite{Yamamoto:2015gzz}, where $C=e^2/(2\pi^2)$ is the 
coefficient of the chiral anomaly. The anomalous transport coefficients $\kappa_B$ and 
$\kappa$ can be expressed from symmetry consideration (parity and charge conjugation 
symmetries) as 
\beq
\label{kappa}
\kappa_B = \tilde \kappa_B e^2 \mu_5, \qquad \kappa = \tilde \kappa T^2,
\eeq
where $\tilde \kappa_B$ and $\tilde \kappa$ are some constants related to the 
coefficients of the chiral anomaly and mixed gauge-gravitational anomaly \cite{Son:2009tf,Landsteiner:2011cp}. 
(The complete expressions of $\kappa$ and $\kappa_B$ can be found, e.g., in 
Ref.~\cite{Landsteiner:2011cp}, but they are unimportant for our purpose in this paper.)
The currents proportional to ${\bm B}$ and ${\bm \omega}$ in Eqs.~(\ref{j}) and (\ref{j_5})
are the CME \cite{Vilenkin:1980fu,Nielsen:1983rb,Alekseev:1998ds,Fukushima:2008xe}
and the chiral vortical effect (CVE) 
\cite{Vilenkin:1979ui,Kharzeev:2007tn,Son:2009tf,Landsteiner:2011cp}, respectively. 

Note that we used the local charge neutrality, $\mu=0$, in Eqs.~(\ref{j}), (\ref{j_5}), and (\ref{kappa})
to ignore the part of the chiral separation effect (CSE) \cite{Son:2004tq, Metlitski:2005pr} 
and the CVE whose transport coefficients include a factor of $\mu$:
${\bm j}^5_{\rm CSE} \propto \mu {\bm B}$, 
${\bm j}_{\rm CVE} \propto \mu \mu_5 {\bm \omega}$, and 
${\bm j}^5_{\rm CVE} \propto \mu^2 {\bm \omega}$. 
We also dropped the contribution of the cross helicity $\propto \mu {\bm v} \cdot {\bm B}$ 
\cite{Yamamoto:2015gzz} in Eq.~(\ref{chiral}). 

Under the local charge neutrality, the displacement current $\d_t{\bm E}$ is negligible,
and Amp\`ere's law becomes  ${\bm j}={\bm \nabla}\times{\bm B}$ \cite{Davidson}. 
Then, Eq.~(\ref{vector}) is automatically satisfied. By eliminating ${\bm j}$ and ${\bm E}$, 
the ChMHD equations for an incompressible fluid (${\bm \nabla} \cdot {\bm v}=0$) above 
reduce to \cite{Yamamoto:2015gzz}
\begin{gather}
\label{v}
(\epsilon+P)(\d_t {\bm v}+{\bm v}\cdot{\bm \nabla}{\bm v})=-\frac{1}{2}{\bm \nabla}{\bm B}^2
+({\bm B}\cdot {\bm \nabla}){\bm B}+\nu{\bm \nabla}^2{\bm v}, \\
\label{B}
\d_t {\bm B}={\bm \nabla}\times({\bm v}\times{\bm B})+\kappa_B{\eta}{\bm \nabla}\times{\bm B}
+\eta{\bm \nabla}^2{\bm B}, \\
\label{n5}
\d_t (n_5 + \kappa {\bm v} \cdot {\bm \omega}) + {\bm v}\cdot{\bm \nabla} n_5 
= -C {\eta} \left[\kappa_B {\bm B}^2 - ({\bm \nabla}\times{\bm B})\cdot{\bm B} \right],
\end{gather}
where $\eta \equiv 1/\sigma$ is the resistivity.
We here ignored the contribution $-{\bm \nabla} P$ on the right-hand side of Eq.~(\ref{v}), 
because, as we will show that $T \gg \mu_5$ in Sec.~\ref{sec:applicability}, the dominant 
contribution to $P$ is the homogeneous $T$.
This set of coupled equations is closed for dynamical variables, 
${\bm v}(t,{\bm x})$, ${\bm B}(t,{\bm x})$, and $\mu_5(t,{\bm x})$.
[As we will discuss in Eq.~(\ref{n_5}) below, $n_5$ is related to $\mu_5$.]
The electric field is given by using these variables as
\beq
{\bm E} = -{\bm v}\times{\bm B} - \kappa_B{\eta} {\bm B}
+ \eta {\bm \nabla}\times{\bm B}.
\eeq

Equations (\ref{v})--(\ref{n5}) can be regarded as an extension of the usual MHD equations 
for relativistic fluids \cite{Biskamp} to the ones with anomalous parity-violating effects 
\cite{Yamamoto:2015gzz}. Indeed, in the absence of anomalous effects 
(setting $n_5=\kappa_B=0$ and disregarding the terms with the coefficient $C$), they 
reduce to the usual MHD equations. These ChMHD equations describe the charged plasmas 
in the early Universe \cite{Joyce:1997uy,Boyarsky:2011uy} and (proto)neutron stars 
\cite{Charbonneau:2009ax,Ohnishi:2014uea} where matter with chirality imbalance 
may be realized. 

\subsection{Regime of applicability}
\label{sec:applicability}
Before proceeding further, we first clarify the regime of the applicability of the ChMHD 
above. It is known that the ChMHD has a plasma instability at finite $\mu_5$ 
\cite{Joyce:1997uy,Boyarsky:2011uy}, called the chiral plasma instability (CPI), whose 
length scale is microscopically estimated as $l_{\rm CPI} \sim (e^2 \mu_5)^{-1}$ 
\cite{Akamatsu:2013pjd}. For the physical picture of the CPI, 
see Ref.~\cite{Akamatsu:2014yza}. 

Recall that hydrodynamics is an effective theory valid at a length scale larger than the 
mean free path. As the mean free path for the $\U(1)$ electromagnetic plasma is given 
by $l_{\rm mfp} \sim (e^4 T)^{-1}$ up to logarithmic corrections, it is necessary to meet 
the following condition for the use of hydrodynamics: $l_{\rm mfp} \ll l_{\rm CPI}$ or 
\beq
\label{condition}
\mu_5(t,{\bm x}) \ll e^2 T.
\eeq
(Otherwise, the ChMHD would have an unstable mode which is beyond its applicability, 
and the theory would not be well defined.)
Under this condition, we have
\beq
\label{n_5}
n_5 \approx \frac{\mu_5 T^2}{6}. 
\eeq
Then, the transport coefficients $\nu$ and $\eta$ can be regarded as constants for 
static and homogeneous $T$. 

It should be remarked that, even in the Maxwell-Chern-Simons equations (or anomalous 
Maxwell equations), which correspond to the limit ${\bm v} \rightarrow {\bm 0}$ of the 
ChMHD, the condition (\ref{condition}) must be satisfied to use the notion of the 
conductivity $\sigma$ itself. This is because $\sigma$ is well defined only at the long 
length scale $l \gg l_{\rm mfp}$. A related point was emphasized in Ref.~\cite{Akamatsu:2013pjd} 
from the viewpoint of the microscopic kinetic theory (see also Ref.~\cite{Akamatsu:2014yza}). 

If one is interested in the physics beyond this regime, $\mu_5 \gtrsim T$, one needs 
to use the chiral kinetic theory \cite{Son:2012wh,Stephanov:2012ki,Chen:2012ca}
instead of the ChMHD, as was done in Ref.~\cite{Akamatsu:2013pjd}. 
This is beyond the scope of the present paper.

\subsection{Conservation laws}
\label{sec:conservation}
In the usual MHD, the magnetic helicity (or the Chern-Simons number),
\beq
{\cal H}_B=\int d^3{\bm x}\, {\bm A}\cdot{\bm B},
\eeq
can be shown to be an approximate conserved quantity for sufficiently large Reynolds 
numbers \cite{Biskamp,Davidson}. On the other hand, one expects that ${\cal H}_B$ 
is not a conserved quantity in the ChMHD due to the CPI. In the following, we consider 
the modifications to the conventional conservation laws.

Using the ChMHD equations above, we obtain the time derivative of the energy $E$ 
and the magnetic helicity ${\cal H}_B$ as
\begin{align}
\label{Et}
\dot E &= - \int d^3{\bm x}\, \left[-\kappa_B \eta {\bm B} \cdot ({\bm \nabla} \times {\bm B}) 
+ \eta ({\bm \nabla} \times {\bm B})^2 \right. \nonumber \\ 
& \qquad \qquad \qquad \left. + \nu (\epsilon + P) ({\bm \nabla} \times {\bm v})^2 \right], 
\\
\label{Ht}
\dot  {\cal H}_B &= 2 \eta \int d^3{\bm x}\,  \left[\kappa_B {\bm B}^2
 - {\bm B} \cdot ({\bm \nabla} \times {\bm B}) \right].
\end{align}
Here the terms with the coefficient $\kappa_B$ on the right-hand sides of Eqs.~(\ref{Et}) 
and (\ref{Ht}) are the modifications to the usual MHD.
As the other terms contain one more derivative compared with the $\kappa_B$ terms
for $B \sim v$, the latter terms are dominant at large length scale, 
$l \gtrsim \kappa_B^{-1} \sim l_{\rm CPI}$. Recalling that $\eta \geq 0$, we have 
$\dot {\cal H}_B \geq 0$ for $\mu_5 > 0$ in this regime. This means that the largest 
change of $E$ and ${\cal H}_B$ occurs at the scale of the CPI, and so we have 
${\dot {\cal H}_B}/{\dot E} \sim l_{\rm CPI}$. Assuming that the integral scale is also 
$l_{\rm CPI}$, we have ${\cal H}_B/E \sim l_{\rm CPI}$. We thus have 
$\dot {\cal H}_B/{\cal H}_B \sim \dot E/E$, and so the magnetic helicity itself is not a 
good conserved quantity except for $\eta = 0$.

This should be contrasted with the conventional MHD, where the change of $E$ and
${\cal H}_B$ occurs only at the scale of dissipation, $l_{\rm mfp}$. In that case, 
$(\dot {\cal H}_B/{\cal H}_B)/(\dot E/E) \sim l_{\rm mfp}/l \ll1$, and ${\cal H}_B$ is 
approximately conserved \cite{Biskamp}. Here $l$ is the scale of turbulence that is 
much larger than $l_{\rm mfp}$ for large Reynolds numbers.

Although the magnetic helicity alone is not conserved, one can show that the total 
helicity, including the helicity of fermions and the helicity of fluids, is conserved. Indeed, 
from Eqs.~(\ref{n5}) and (\ref{Ht}) [or directly from Eq.~(\ref{chiral})], we obtain 
the conservation of total helicity \cite{Yamamoto:2015gzz},
\beq
\d_t {\cal H}_{\rm tot}=0, \qquad 
{\cal H}_{\rm tot} \equiv \frac{C}{2}{\cal H}_B + {\cal H}_v + N_5\,,
\eeq
where
\beq
N_5 \equiv \int d^3{\bm x}\, n_5, \qquad {\cal H}_v \equiv \int d^3{\bm x}\, \kappa {\bm v} \cdot {\bm \omega}
\eeq
are the helicity (or chiral charge) of fermions and the fluid helicity, respectively. 
Note here that the cross helicity $\propto \int \mu {\bm v} \cdot {\bm B}$ \cite{Yamamoto:2015gzz}
is absent under the local charge neutrality, $\mu=0$.

\subsection{Scaling symmetry}
\label{sec:scaling}
We now turn to the scaling symmetry of the ChMHD. Let us first recall the scaling symmetry 
of the usual MHD. For $n_5=\kappa_B=0$, ignoring the $C$ terms, Eqs.~(\ref{v}) and (\ref{B}) 
are invariant under the scaling \cite{Olesen:1996ts}
\begin{gather}
{\bm x} \rightarrow l{\bm x}, \quad t \rightarrow l^{1-h}t, \quad {\bm v} \rightarrow l^{h}{\bm v}, 
\quad {\bm B} \rightarrow l^{h}{\bm B}, \nonumber \\
\nu \rightarrow l^{1+h} \nu, \quad \eta \rightarrow l^{1+h} \eta,
\label{symmetry}
\end{gather}
where $l$ is the positive scaling factor and $h$ is any real parameter. 
The transformation laws for other variables follow from Maxwell's equations (\ref{Maxwell})
as, e.g., ${\bm E} \rightarrow l^{2h}{\bm E}$. Imposing the condition that the coefficients 
$\nu$ and $\eta$ are nonzero constants, $h$ is fixed as a unique value, $h=-1$. On the 
other hand, in the inertial range where the $\nu$ and $\eta$ terms are negligible, the MHD 
has generic scaling symmetries with {\it any} $h$ \cite{Olesen:1996ts}.

It is easy to check that ChMHD equations (\ref{v})--(\ref{n5}) have the same scaling 
symmetry in the absence of the local fluid helicity (${\bm v} \cdot {\bm \omega} = 0$) if 
we further impose the following scaling for $\mu_5$ and $n_5$ at the same time%
\footnote{A partial transformation law (\ref{symmetry}), which does not take into account 
the anomaly relation (\ref{n5}) and the scaling (\ref{mu5}), was previously given in 
Ref.~\cite{Giovannini:2013oga}.}:
\beq
\label{mu5}
\mu_5 \rightarrow l^{-1}\mu_5, \quad n_5 \rightarrow l^{1+2h}n_5.
\eeq
As $\mu_5$ and $n_5$ are related by $n_5 \propto \mu_5$ for $\mu_5 \ll T$ 
as shown in Eq.~(\ref{n_5}), $h$ is fixed as 
\beq
\label{h}
h = -1.
\eeq
Coincidentally, this is the same value as the one required by constant $\nu$ and $\eta$. 

It should be remarked that the local charge neutrality and the absence of the local fluid helicity 
are essential for this scaling symmetry; if $\mu \neq 0$, we would have the CVE of the form 
${\bm j} \propto \mu \mu_5 {\bm \omega}$ in Eq.~(\ref{j}), which would violate the scaling 
symmetry above. The presence of the local fluid helicity $\kappa {\bm v} \cdot {\bm \omega}$
would also break down the scaling symmetry.

In the inertial range, the chiral anomaly with the coefficient $C$ and the CME with the coefficient 
$\kappa_B$ do not contribute at all in Eqs.~(\ref{v}) and (\ref{B}), while the local fluid helicity 
$\kappa {\bm v} \cdot {\bm \omega}$ can. In this case, the ChMHD has generic scaling 
symmetries with {\it any} $h$, even in the presence of the fluid helicity, if we impose the 
following scaling for $n_5$:
\beq
n_5 \rightarrow l^{-1+2h} n_5.
\eeq

\subsection{Physical consequences}
\label{sec:consequence}
Let us explore the physical consequences of the scaling symmetry (\ref{mu5}) 
with $h=-1$ in the turbulent regime. Our argument here is analogous to the one 
in Refs.~\cite{Olesen:1996ts, Campanelli:2004wm}. 
We will first leave $h$ unspecified for later purposes and will set $h=-1$ later.

We first define the average chiral density,
\beq
\bar n_5(t)\equiv \frac{1}{V}\int d^3{\bm x}\, n_5({\bm x}, t)\,,
\eeq
where $V=\int^L_{2\pi/K}d^3{\bm x}$, with $2\pi/L$ and $K$ being the infrared
and ultraviolet momentum cutoffs, respectively. In the following, we will consider
the formal limit as $L \rightarrow \infty$ and $K \rightarrow \infty$.
From Eq.~(\ref{chiral}), and assuming that ${\bm j}_5$ vanishes at sufficiently 
large distances, the time evolution of $\bar n_5(t)$ is given by
\beq
\label{n5_bar}
\d_t \bar n_5=\frac{C}{V}\int d^3{\bm x}\, {\bm E}\cdot{\bm B}
=\int_{0}^{\infty} dk\,{\cal N}(k,t)\,,
\eeq
where
\beq
\label{N}
{\cal N}(k,t)=\frac{4\pi C}{V}k^2 \langle {\bm E}({\bm k}, t)\cdot{\bm B}^*({\bm k}, t) \rangle \,,
\eeq
for an isotropic turbulence.

We also consider the magnetic and kinetic energy densities in $k \equiv |{\bm k}|$ space:
\begin{align}
\label{E_B}
{\cal E}_B(k, t) &= \frac{2\pi k^2}{(2\pi)^3} \int d^3 {\bm y} \ e^{i {\bm k} \cdot {\bm y}}
\langle {\bm B}({\bm x}, t) \cdot {\bm B}({\bm x} + {\bm y}, t) \rangle \,, \\
\label{E_v}
{\cal E}_v(k, t) &= \frac{2\pi k^2}{(2\pi)^3} \int d^3 {\bm y} \ e^{i {\bm k} \cdot {\bm y}}
\langle {\bm v}({\bm x}, t) \cdot {\bm v}({\bm x} + {\bm y}, t) \rangle \,, \nonumber \\
\end{align}
and the magnetic and kinetic correlation lengths defined by
\begin{align}
\label{xi_B}
\xi_B (t) &= 2\pi \frac{\int_0^{\infty} dk \ k^{-1}{\cal E}_B(k,t)}{\int_0^{\infty} dk \ {\cal E}_B(k,t)}\,, \\
\label{xi_v}
\xi_v (t) &= 2\pi \frac{\int_0^{\infty} dk \ k^{-1}{\cal E}_v(k,t)}{\int_0^{\infty} dk \ {\cal E}_v(k,t)}\,,
\end{align}
respectively.

Let us now look into the scaling symmetries of ${\cal N}(k,t)$, ${\cal E}_B(k, t)$, and 
${\cal E}_v(k, t)$. From Eq.~(\ref{symmetry}), they satisfy
\begin{gather}
\label{scaling}
{\cal N}(l^{-1}k,l^{1-h}t) = l^{1+3h}{\cal N}(k,t), \\
{\cal E}_B(l^{-1}k, l^{1-h}t) = l^{1+2h}{\cal E}_B(k,t), \\
{\cal E}_v(l^{-1}k, l^{1-h}t) = l^{1+2h} {\cal E}_v(k, t). 
\end{gather}
We introduce the functions $\psi_n(k,t) \equiv k^{1+3h}{\cal N}(k,t)$, 
$\psi_B(k,t) \equiv k^{1+2h}{\cal E}_B(k,t)$, and 
$\psi_v(k,t) \equiv k^{1+2h}{\cal E}_v(k,t)$,
such that
\begin{gather}
\psi_n(l^{-1}k,l^{1-h}t)=\psi_n(k,t), \\
\psi_B(l^{-1}k,l^{1-h}t)=\psi_B(k,t), \\
\psi_v(l^{-1}k,l^{1-h}t)=\psi_v(k,t).
\end{gather}
These relations mean that $\psi_n$, $\psi_B$, and $\psi_v$ are functions of $x \equiv k^{1-h}t$ alone:
$\psi_n(k,t)=\psi_n(k^{1-h}t)$, $\psi_B(k,t)=\psi_B(k^{1-h}t)$, and $\psi_v(k,t)=\psi_v(k^{1-h}t)$. 
Hence, ${\cal N}(k,t)$, ${\cal E}_B(k, t)$, and  ${\cal E}_v(k, t)$ can be expressed as
\beq
\label{N-psi}
{\cal N}(k,t) &=k^{-1-3h}\psi_n(k^{1-h}t), \\
\label{E_B-psi}
{\cal E}_B(k, t) &=k^{-1-2h}\psi_B(k^{1-h}t), \\
\label{E_v-psi}
{\cal E}_v(k, t) &=k^{-1-2h}\psi_v(k^{1-h}t).
\eeq

Substituting Eqs.~(\ref{N-psi}), (\ref{E_B-psi}), and (\ref{E_v-psi}) into 
Eqs.~(\ref{n5_bar}), (\ref{xi_B}), and (\ref{xi_v}), respectively, and performing 
the integral over $t$ with assuming $\bar n_5(\infty)=0$ in the first,%
\footnote{As seen from Eq.~(\ref{n_5_scaling}), this assumption can be satisfied 
when $h < -\frac{1}{2}$ or $h>1$.}
we have
\begin{align}
\label{n_5_scaling}
\bar n_5(t)&=\bar n_5(t_s)\left(\frac{t}{t_s}\right)^{\! \! \frac{1+2h}{1-h}}\,, \\
\label{xi_B_scaling}
\xi_B(t)&=\xi_B(t_s) \left(\frac{t}{t_s} \right)^{\! \! \frac{1}{1-h}}\,, \\
\label{xi_v_scaling}
\xi_v(t)&=\xi_v(t_s) \left(\frac{t}{t_s} \right)^{\! \! \frac{1}{1-h}}\,,
\end{align}
where $t_s$ is some parameter and
\begin{align}
\bar n_5(t_s) &= \frac{1}{1+2h}t_s^{\frac{1+2h}{1-h}}\int_0^{\infty}dx \ x^{-\frac{1+2h}{1-h}}\psi_n(x)\,, \\
\xi_B(t_s) &= 2\pi t_s^{\frac{1}{1-h}} \frac{\int_0^{\infty} dx \ x^{-\frac{2+h}{1-h}} \psi_B(x)}{\int_0^{\infty} dx \ x^{-\frac{1+h}{1-h}} \psi_B(x)}\,, \\
\xi_v(t_s) &= 2\pi t_s^{\frac{1}{1-h}} \frac{\int_0^{\infty} dx \ x^{-\frac{2+h}{1-h}} \psi_v(x)}{\int_0^{\infty} dx \ x^{-\frac{1+h}{1-h}} \psi_v(x)}\,.
\end{align}

Inserting $h=-1$, corresponding to the unique scaling symmetry (\ref{h}) in the ChMHD, 
we obtain
\begin{align}
\label{n_5_t}
\bar n_5(t) &= \bar n_5(t_s)\left(\frac{t_s}{t}\right)^{\! \! \frac{1}{2}}\,, \\
\label{xi_B_t}
\xi_B(t) &=\xi_B(t_s) \left(\frac{t}{t_s} \right)^{\! \! \frac{1}{2}}\,,\\
\label{xi_v_t}
\xi_v(t) &=\xi_v(t_s) \left(\frac{t}{t_s} \right)^{\! \! \frac{1}{2}}\,.
\end{align}
We expect that the solutions of the ChMHD asymptotically approach these behaviors 
regardless of the initial conditions. In particular, Eq.~(\ref{xi_B_t}) and (\ref{xi_v_t}) 
show that both $\xi_B(t)$ and $\xi_v(t)$ grow with time, meaning that both the magnetic and 
kinetic energies are transferred from a small scale to a larger scale: the inverse energy cascade.

Equation (\ref{N-psi}) with $h=-1$ exhibits the same self-similar inverse cascade of magnetic 
helicity observed in the Maxwell-Chern-Simons theory \cite{Hirono:2015rla}. Our result here 
provides its generalization to the ChMHD, together with the new result (\ref{xi_v_t}), even in 
the presence of the fluid velocity ${\bm v}$. This argument shows that the self-similar behaviors 
in Eqs.~(\ref{N-psi})--(\ref{E_v-psi}) with $h=-1$ can be seen as a consequence of the scaling 
symmetry (\ref{symmetry}) and (\ref{mu5}) in the ChMHD. However, it would break down 
away from the charge neutrality or in the presence of fluid helicity, as we have seen above.

\section{Chiral hydrodynamics for neutral matter}
\label{sec:ChHD}
We then consider the chiral hydrodynamics for neutral matter of a \emph{single} chiral 
fermion at finite chemical potential $\mu \neq 0$. Our primary interest here is the 
application to the neutrino hydrodynamics considered in Ref.~\cite{Yamamoto:2015gzz}.

\subsection{Hydrodynamic equations}
\label{sec:equation_n}
As neutral matter does not couple to electromagnetic fields, the hydrodynamic equation 
in this case is 
\beq
\label{v_n}
(\epsilon + P)(\d_t + {\bm v} \cdot {\bm \nabla}) {\bm v}
= - {\bm \nabla} P + \nu {\bm \nabla}^2 {\bm v}.
\eeq
This is the usual relativistic hydrodynamics to the leading order in $v$ \cite{Landau}. 
We here include the contribution $- {\bm \nabla} P$ unlike Eq.~(\ref{v}) in the ChMHD,
because we can consider not only the regime $\mu \ll T$, but also $\mu \gg T$ in the
neutral chiral hydrodynamics, where $\mu$ is generally inhomogeneous (see below). 
On the other hand, the current conservation is modified by the CVE as \cite{Yamamoto:2015gzz}
\begin{gather}
\label{n_n}
\d_t \left(n + \kappa {\bm v} \cdot {\bm \omega} \right) + {\bm \nabla} \cdot {\bm j} = 0, \\ 
\label{j_n}
{\bm j} = n {\bm v} + \kappa {\bm \omega},
\end{gather}
where $\kappa = \tilde \kappa_1 \mu^2 + \tilde \kappa_2 T^2$ with some constants 
$\tilde \kappa_{1,2}$ (see Ref.~\cite{Landsteiner:2011cp} for the detailed expressions).

Note that the neutral chiral matter does not have the CPI, unlike the charged chiral 
plasmas in Sec.~\ref{sec:ChMHD}. Hence, we do not have the constraint like 
Eq.~(\ref{condition}) in the present case.

\subsection{Scaling symmetry}
\label{sec:scaling_n}
Let us now consider the scaling symmetry of the chiral hydrodynamics for neutral matter 
above. First, when $\epsilon$ and $P$ are constants, Eq.~(\ref{v_n}) has the following 
scaling symmetry:
\begin{gather}
\label{symmetry_n}
{\bm x} \rightarrow l{\bm x}, \quad t \rightarrow l^{1-h}t, \quad {\bm v} \rightarrow l^{h}{\bm v}, 
\quad \nu \rightarrow l^{1+h} \nu, 
\end{gather}
for any $h$. However, once the conservation law (\ref{n_n}) is taken into account,
this scaling symmetry seems not to hold for any $h$, even in the inertial range, at first sight. 

In fact, there is a regime where the hydrodynamic equations above have a scaling symmetry.
The point here is that, despite the absence of the CPI, the number density $n$ can vary due 
to the CVE in Eq.~(\ref{n_n}) \cite{Yamamoto:2015gzz}, so that $n$ must be regarded as a 
dynamical variable, $n(t,{\bm x})$. We thus impose the scaling for $\mu$ as
\beq
\label{mu_n}
\mu \rightarrow l^p \mu,
\eeq
with some real parameter $p$. 

We now show that the chiral hydrodynamics has a unique scaling symmetry,
\beq
\label{hp}
h=0, \qquad p=-1,
\eeq
both for $\mu \ll T$ and $\mu \gg T$ in the inertial range  where the $\nu$ term can be ignored.

When $\mu \ll T$, the thermodynamic quantities and the transport coefficient $\kappa$ 
depend on $T$ and $\mu$ as $\epsilon \propto T^4$, $P \propto T^4$, $n \propto \mu T^2$,
and $\kappa \propto T^2$ to the leading order in $\mu/T \ll 1$. For Eqs.~(\ref{n_n}) and 
(\ref{j_n}) to possess a scaling symmetry, $n \sim \kappa {\bm v} \cdot {\bm \omega}$ and 
$n {\bm v} \sim \kappa {\bm \omega}$ (where ``$\sim$" stands for the same scaling exponent),
we must have 
\beq
p = 2h - 1, \qquad h + p = h - 1.
\eeq
The solution of these equations is given by Eq.~(\ref{hp}).
Then it is easy to check that Eq.~(\ref{v_n}) satisfies this scaling symmetry in the inertial 
range where the dissipative term $\nu$ can be ignored.

When $\mu \gg T$, on the other hand, we have $\epsilon \propto \mu^4$, $P \propto \mu^4$, 
$n \propto \mu^3$, and $\kappa \propto \mu^2$ to the leading order in $T/\mu \ll 1$. 
Imposing a scaling symmetry in Eqs.~(\ref{n_n}) and (\ref{j_n}), we must have 
\beq
3p = 2h + 2p - 1, \qquad h + 3p = h + 2p - 1,
\eeq
leading to Eq.~(\ref{hp}) again. Similarly to above, Eq.~(\ref{v_n}) satisfies this scaling symmetry 
in the inertial range.

In summary, the neutral chiral hydrodynamics in the inertial range has the same scaling 
symmetry (\ref{symmetry_n}) and (\ref{mu_n}) with $h$ and $p$ given by Eq.~(\ref{hp}) 
both when $\mu \ll T$ and when $\mu \gg T$. The exponent $h$ in this case is uniquely 
determined by the presence of the CVE, but it is different from Eq.~(\ref{h}) in the ChMHD. 
This uniqueness should be contrasted with the generic scaling symmetries of the usual 
hydrodynamics with any $h$ in the inertial range. Note, however, that this unique scaling 
symmetry is lost outside the inertial range.

\subsection{Physical consequences}
\label{sec:consequence_n}
Let us study the physical consequences of the scaling symmetry (\ref{symmetry_n}) 
and (\ref{mu_n}) in the chiral hydrodynamics for neutral matter in the turbulent regime
where the kinetic Reynolds number is sufficiently large. We consider the kinetic energy 
density defined in Eq.~(\ref{E_v}) and the kinetic correlation length in Eq.~(\ref{xi_v}).

From the scaling symmetry (\ref{symmetry_n}), it follows that
\beq
{\cal E}_v(l^{-1}k, l^{1-h}t) = l^{1+2h} {\cal E}_v(k, t). 
\eeq
Then, we can use the same argument in Sec.~\ref{sec:consequence}, leading to
Eq.~(\ref{E_v-psi}) for the kinetic energy density and Eq.~(\ref{xi_v_scaling}) for the
kinetic correlation length.

Inserting $h=0$ as required by Eq.~(\ref{hp}) in the neutral chiral hydrodynamics, 
we arrive at
\begin{gather}
\label{E_v_n}
{\cal E}_v(k, t) = k^{-1}\psi_v(kt), \\
\label{xi_v_t_n}
\xi_v(t) = \xi_v(t_s) \left(\frac{t}{t_s} \right)\,,
\end{gather}
which we expect to hold universally at late times. Equation~(\ref{xi_v_t_n}) shows the 
inverse energy cascade. Note here that the time dependence of $\xi_v(t)$ in Eq.~(\ref{xi_v_t_n}) 
is different from that of $\xi_v(t)$ in Eq.~(\ref{xi_v_t}) in the ChMHD; $\xi_v(t)$ in the neutral 
chiral hydrodynamics grows faster than $\xi_v(t)$ in the ChMHD because of the different 
scaling symmetries between Eqs.~(\ref{h}) and (\ref{hp}).

\section{Summary and discussion}
\label{sec:discussion}
In this paper, we found that the chiral (magneto)hydrodynamic equations for charged and 
neutral matter in the turbulent regime have {\it unique} scaling symmetry under certain 
conditions. These scaling symmetries dictate the behaviors of the chiral charge and magnetic 
and kinetic correlation lengths: $n_5(t) \sim t^{-1/2}$ and $\xi_B(t) \sim \xi_v(t) \sim t^{1/2}$ 
in charged chiral matter and $\xi_v(t) \sim t$ in neutral chiral matter 
[see Eqs.~(\ref{n_5_t})--(\ref{xi_v_t}) and (\ref{xi_v_t_n})]. These scaling laws suggest the 
inverse energy cascade in both charged and neutral chiral matter.

Among others, our results may have potential relevance in core-collapse supernovae, 
where the chiral transport of neutrinos are expected to play key roles \cite{Yamamoto:2015gzz}.
Since their dynamical evolution is described by the coupled transport equations for neutrinos, 
electrons, and baryons, the simple scaling symmetries and scaling laws derived here may 
not be directly applicable. Nonetheless, the fact that the inverse energy cascade occurs 
both in the ChMHD and in neutral chiral hydrodynamics suggests the tendency toward the 
inverse energy cascade in the presence of the chiral transport of neutrinos. If this is the case, 
it should work favorably for supernova explosions compared with the direct energy cascade 
observed in the conventional neutrino transport without the effects of chirality or helicity 
\cite{Hanke:2011jf}. It should be important to check the possible inverse cascade numerically 
by the future three-dimensional {\it chiral} neutrino radiation hydrodynamics.

For quark-gluon plasmas created in heavy ion collisions, the bulk fluid motion is relativistic.
In this case, because of the $\gamma$ factor in relativistic hydrodynamics, 
$\gamma = 1/\sqrt{1 - {\bm v}^2}$, there is no scaling symmetry like Eqs.~(\ref{symmetry}) 
and (\ref{mu5}), and the self-similar behaviors and scaling laws like Eqs.~(\ref{n_5_t}), 
(\ref{xi_B_t}), and (\ref{xi_v_t}) are lost. The fate of the ChMHD turbulence in 
these ultrarelativistic systems would be an interesting question to be investigated in the future.

While we have concentrated on the self-similarity of the chiral (magneto)hydrodynamics in 
this paper, it would be interesting to study the possible self-similarity at the level of the 
chiral kinetic theory. Finally, one can also ask the possible effects of finite fermion masses 
and nonlinear chiral transport phenomena \cite{Chen:2016xtg, Gorbar:2016qfh} on the 
scaling laws in the turbulent regime.

\acknowledgments
We thank Y.~Akamatsu for useful discussions. 
This work was supported, in part, by JSPS KAKENHI Grant No.~26887032
and the MEXT-Supported Program for the Strategic Research Foundation
at Private Universities, ``Topological Science" (Grant No.~S1511006).

\end{document}